\documentclass[twocolumn]{aa} 
\usepackage{graphicx}
\begin{document}

  \title{Evolution from AGB to planetary nebula in the MSX survey
     \thanks{Based on results obtained by the MSX survey}
     \thanks{Tables 1 to 3 are only available in electronic form at the CDS
via anonymous ftp to cdsarc.u-strasbg.fr (130.79.128.5) or via
http://cdsweb.u-strasbg.fr/cgi-bin/qcat?J/A+A/}}


   \author{R. Ortiz\inst{1}, S. Lorenz-Martins\inst{2}, W.J. Maciel\inst{3},
		E.M. Rangel\inst{1}}
		
   \institute{Departamento de F\'\i sica, UFES, 
              	Av. Fernando Ferrari 514, 29075-910, 
		Vit\'oria, ES, Brazil \\
              \email{ortiz@cce.ufes.br}
         \and
        	GEMAC, Depto. de Astronomia, OV/UFRJ,
		Ladeira Pedro Antonio 43, 20080-090,
		Rio de Janeiro, RJ, Brazil\\
            \email{slorenz@ov.ufrj.br} 
        \and 
             	Departamento de Astronomia, IAG/USP, 
		Rua do Mat\~ao, 1226, Cidade Universit\'aria, 05508-900, 
		S\~ao Paulo, SP, Brazil
            \email{maciel@astro.iag.usp.br}
		}

\authorrunning{R.Ortiz et al.}
\titlerunning{AGB to planetary nebula evolution} 
 
   \offprints{R.Ortiz}

   \date{Received; accepted}

   \abstract{
	We investigate the evolution of oxygen- and carbon-rich AGB stars,
post-AGB objects, and planetary nebulae using data collected mainly from
the MSX catalogue. Magnitudes and colour indices are compared with those
calculated from a grid of synthetic spectra that describe the post-AGB
evolution beginning at the onset of the superwind. We find that carbon
stars and OH/IR objects form two distinct sequences in the (K$-$[8.3])
$\times$([8.3]$-$[14.7]) MSX colour diagram. OH/IR objects are distributed
in two groups: the bluest ones are crowded near [14.7]$-$[21.3]$\simeq 1$
and [8.3]$-$[14.7]$\simeq 2$, and a second, redder group is spread over
a large area in the diagram, where post-AGB objects and planetary nebulae
are also found. High mass-loss rate OH/IR objects, post-AGB stars, and
planetary nebulae share the same region in the (K$-$[8.3])$\times$
([8.3]$-$[14.7]) and [14.7]$-$[21.3]$\times$([8.3]$-$[14.7]) colour-colour
diagrams. This region in the diagram is clearly separated from a bluer one
where most OH/IR stars are found. We use a grid of models of post-AGB
evolution, which are compared with the data. The gap in the colour-colour
diagrams is interpreted as the result of the rapid trajectory in the
diagram of the stars that have just left the AGB.

   }

\maketitle
%

\section{Introduction}

AGB and post-AGB evolution have lately become better understood,
thanks to numerous surveys carried out by different groups
at several wavelengths, from radio (e.g. CO rotational transitions,
OH maser emission at 18 cm) to infrared. The present knowledge of
stellar evolution in the AGB (Iben Jr., 1995) states that the more
massive stars in this phase may undergo a third dredge-up episode,
which enriches their convective envelope with carbon that will
eventually reach the atmosphere resulting in a ``carbon-rich star''.
However, if the star does not experience the third dredge-up, or if
its evolution has not reached this point, its atmosphere remains
oxygen-rich. Therefore the chemical composition of the atmosphere
in the AGB is dependent on both the initial mass {\it and}
the evolutionary phase of the star. In any case, when the mass loss
rate increases, which is believed to happen at the end of the AGB,
an optically thick circumstellar dust shell (CDS) develops, making
the star invisible at visual wavelengths. The star then climbs up
the AGB becoming very luminous
($L_{\rm bol} \simeq 10^3 - 10^6 L_{\odot}$)
especially in the infrared, most of its flux being emitted by cold and
warm grains ($T_{\rm dust} \sim 300 - 900$ K, depending on the chemical
composition) in the CDS. This twofold scenario (large infrared flux
and low colour temperature) makes AGB stars, together with young stellar
objects, the main body of detections in infrared surveys. They are
detected at large distances, such as several kiloparsecs, as far as the
galactic bulge and beyond. In addition to infrared continuum
emission, AGB stars may exhibit a variety of
molecular lines including CO, HCN (in carbon-rich); OH, SiO, and H$_2$O
(in oxygen-rich stars). The emission originates in the circumstellar
shell and, together with some features observed in the infrared, it
represents the chemical (C/O) signature of the atmosphere of the star.
Hydroxyl maser emission at 18 cm is quite common and its 1612 MHz component
becomes particularly strong in optically thick shells; this is widely
known as the OH/IR phenomenon.

After the AGB phase the star follows a horizontal track in the
Hertzprung-Russell diagram (Sch\"onberner 1981, 1983). The convective
envelope of the star is expelled and its hot core revealed. This
process takes about $10^3 \sim 10^4$ years before the luminosity of the
core decreases and the star moves towards the white dwarf cooling track.
Immediately after the envelope is expelled a reflection nebula appears
around the star.
Eventually a small HII region appears in the inner part of the nebula
followed by metallic emission lines of increasing ionization potential in the
spectrum. This hybrid object, which still shows some characteristics of
the AGB phase like intense infrared continuum and molecular bands, is
often called ``post-AGB'' star or ``proto planetary nebula''. Here we
prefer the name ``transition object'', without making distinctions in
the true evolutionary status during its horizontal trajectory after
the end of the AGB. 
The study of these objects has in fact shown that they are evolved
objects, with chemical abundances and metallicities that resemble those
of AGB stars, showing enhanced s-process and CN abundances. About one
to two hundred candidates have been suggested, observed and confirmed
as transition objects, so far.

Van der Veen \& Habing (1988, hereafter VH) studied the distribution of
different kinds of AGB star in the [12-25]$\times$[25-60] colour-colour
diagram. The diagram was divided into regions, each corresponding to one
type of object. Regions I through IIIb (see Fig. 5b in VH) contain the
stars evolving from O-rich Miras to OH/IR stars. Region IV mainly
consists of transition objects, and region V of planetary nebulae.
The release of the MSX point source catalogue (Egan et al. 1997) has
offered a new opportunity to study the AGB to PN evolution in mid-infrared
colour-colour diagrams. The main scope of this work is to verify whether
the AGB to PN evolution in MSX colour-colour diagrams follows a pattern
similar to what was found by VH using the IRAS colour-colour diagram.
We make use of samples containing OH/IR objects, C-rich AGB stars,
transition objects, and planetary nebulae, to study how these objects
are spread in different MSX (and near-infrared) colour-colour diagrams.
We also compare the distributions above with predictions of synthetic
theoretical models of post-AGB evolution. The scheme of this paper is
as follows: in Sect. 2 we present an overview of the database used
in this work; in Sect. 3 colour-colour diagrams are discussed;
Sect. 4 contains a discussion about the nature of the transition
objects; Sect. 5 is devoted to the comparison between theoretical
models for post-AGB evolution and the data; and in section 6 we
present our conclusions.

\section{Surveys of evolved objects: the data}

The major mid-infrared database used in this work is the point source
catalogue of the MSX (Midcourse Space Experiment; Egan et al. 1997) project.
The main scope of the MSX survey was to cover the whole galactic plane
at $\vert b \vert < 5^o$ in the infrared, between $4.2 \mu$m
$ < \lambda < 27 \mu$m. This region had been formerly covered by the
IRAS satellite, however due to various technical limitations, the IRAS
mission did not have spatial resolution suited to the crowdedness of
the galactic plane. On the other hand, MSX combines better spatial resolution
($~18.3$ arc seconds), with sensitivity and wide wavelength coverage which
allows one to resolve red giant stars in moderately crowded regions in the
galactic plane.
Version 1.0 of the catalogue, which is used in this work, contains
323,052 sources. All MSX flux densities have been converted into
magnitudes according to the Explanatory Guide (Egan et al. 1999).
The zero-magnitude flux densities are the following: 58.49, 26.51, 18.29,
and 8.80 Jansky, corresponding to bands A, C, D, and E, respectively.
These zero-points were calculated relative to a blackbody with $10,000$ K
temperature. The isophotal wavelengths of these bands are: $8.28 \mu$m
(band A), $12.13 \mu$m (band C), $14.65 \mu$m (band D), and $21.34 \mu$m
(band E). Bands B1 and B2, centered near $\lambda = 4.3 \mu$m were
discarded in this work due to their poor sensitivity.

A large number of molecular surveys has been carried out towards several
regions of the Galaxy. OH surveys are especially suited to detect OH/IR
stars. Benson et al. (1990) compiled a catalogue of OH observations made
by several authors. More recently, OH observations have been carried out
towards the bulge (Sjouwerman 1998, Lindqvist 1998) and the galactic plane
(Sevenster 1997a,b; 2001), among others. Carbon-rich stars can be detected at
microwave frequencies, especially by observing the CO and HCN transitions.
In this work we use the list of microwave observations of these two
molecules made by Loup et al. (1993), comprising 213 stars identified
in the literature as carbon-rich stars. The OH/IR objects are
taken from the galactic plane OH maser surveys of Sevenster cited above,
amounting to 766 objects. The sample of transition objects
comprises 106 objects taken from several papers in the literature,
consisting of a wide range of evolutionary stages, ranging from AGB stars
showing evidence of a superwind to newborn planetary nebulae. The catalogue
of planetary nebulae used in this study is that of Kimeswenger
(2001), which contains 995 objects in the southern hemisphere.

The sample of carbon stars mentioned above (Loup et al. 1993) was
cross-correlated with the MSX catalogue of point sources. One must keep
in mind that the compilation of Loup et al. contains stars distributed
in all directions, observed with different radio telescopes and thus the
spatial resolution changes from source to source. We chose to search the
MSX catalogue within 10 arc seconds from the radio positions as Loup et al.
consider this figure representative for objects that match an IRAS source.
This is much larger than the astrometric accuracy claimed by the MSX team
($\sim 2$ arc seconds) but still smaller than the spatial resolution of
the MSX survey (18.3''). Near-infrared counterparts were mainly taken from
the literature (Guglielmo et al. 1993, 1997, 1998; Gezari et al. 1987).
These papers provide the $K-L'$ colour, which is an excellent estimator
of the optical depth of the circumstellar shell (L\'epine et al. 1995,
Epchtein et al. 1987). When the near-infrared photometry was not found
in the references above, we looked for a 2MASS counterpart within 3
arc seconds of the radio position. The search above resulted in 45
MSX counterparts (Table 1), all of them with a counterpart in the
near-infrared, taken either from the literature or from the 2MASS
point-source catalogue.

The list of OH maser sources by Sevenster (1997a,b; 2001) was obtained
using both the ATCA and the VLA, which makes these data more homogeneous
concerning astrometric accuracy and flux calibration, for instance, than
the present sample of carbon stars.
Infrared counterparts of the 766 objects detected in these OH surveys
have been studied by Sevenster (2002a,b), who found 494 MSX and 194 
2MASS objects within 5'' and 3'' of the radio position, respectively.
In this work, when cross-correlation involving OH, MSX and 2MASS is
needed, we assume the identifications provided by Sevenster (2002a).
To the 194 2MASS counterparts identified by Sevenster we added 28
more, taken from various sources in the literature, amounting to
a total of 222 objects.

Since we intend to study how AGB stars evolve into planetary nebulae, it
is important to compile a sample that contains a wide range of
characteristics, especially the optical depth of the envelope, which
depends mainly on the mass-loss rate. In this case, near-infrared colour
indices are more appropriate than indices involving the near-and-mid-infrared,
because the latter are often influenced by numerous molecular features
like SiO. Figure 1 shows the $(J-H) \times (H-K)$ colour-colour diagram
for carbon-rich stars and OH/IR objects that have a MSX counterpart.
One can see that the present sample includes stars in a wide range of
colours up to $J-H=4.6$. Therefore we can assume that our sample also
contains stars in a wide range of optical depths and mass-loss rates.

The present sample of {\it transition objects} amounts to 106 stars, all
of them having an IRAS counterpart and $\vert b \vert < 5^o$, which is the
area surveyed by MSX. It was compiled from numerous sources in the
literature. Similarly to the carbon stars, we searched for MSX
counterparts within 10'' of the IRAS position and as a result we obtained
69 sources (Table 2). Most of them (64 objects $= 93$ \%) have a
near-infrared counterpart either in the literature or taken from the 2MASS
database.

\begin{figure}
\includegraphics[width=9cm]{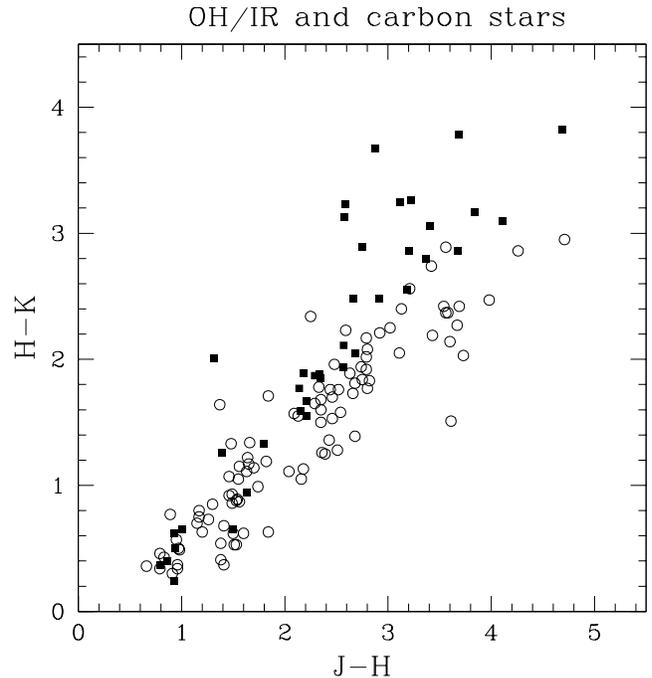}
\caption[]{$(J-H) \times (H-K)$ colour-colour diagram of C-rich
(squares) and OH/IR objects (open circles), with a MSX counterpart.
The near-infrared photometry was extracted from the literature and
the 2MASS database.}
\end{figure}

Among the 995 planetary nebulae in the list of Kimeswenger, 606 have
$\vert b \vert < 5^o$, the area covered by MSX. We searched for MSX
counterparts within 10'' of the catalogue position, and found 214
objects. Among these, 162 have diameters $\theta < 18$'', which is the
spatial resolution of the MSX survey, listed in the catalogue by Acker
et al. (1992). These nebulae can be considered as point sources and have
been selected for study (Table 3). Among these 214 objects, 89 nebulae
have reliable K magnitudes, whereas 73 have their magnitudes listed as
lower limits.

Table 4 collects the figures concerning the several source catalogues used
in this work, their sky coverage, number of MSX counterparts found,
near-infrared counterparts, etc. All objects selected for study in
this paper have an MSX counterpart found according to the criteria described
above.

\setcounter{table}{3}
\begin{table*}
\caption[]{Summary of the database. The {\it number of objects} in the last
three columns refers to: (1) in the original reference; (2) with an MSX
counterpart; (3) with an MSX {\it and} near-infrared counterpart. All
objects selected for study in this paper have a counterpart in the MSX
survey, which covered the whole galactic plane restricted to 
$\vert b \vert < 5^o$.
}
\begin{tabular}{|c|c|c|ccc|}
\hline
\hline
\noalign{\smallskip}
Object & Reference to the & sky coverage & \multicolumn{3}{|c|}
{Number of objects} \\
 & original database & & (1) & (2) & (3) \\

\noalign{\smallskip}
\hline
OH/IR & Sevenster 1997a,b; 2001 & $\vert l \vert < 45^o, \vert b \vert < 3^o$
& 766 & 494 & 222 \\

C-rich stars & Loup et al. 1993 & all-sky &
213 & 45 & 45 \\

transition objects & several & $\vert b \vert < 5^o$ &
106 & 69 & 64 \\

planetary nebulae & Kimeswenger 2001 & $\delta < +2^o$ & 
995 & 214 & 86 \\

\noalign{\smallskip}
\hline
\hline
\end{tabular}
\end{table*}

\section{Post-AGB evolution in colour-colour diagrams}

Colour-colour diagrams have often been used to discriminate among
different classes of AGB stars, especially concerning their chemical
features. After the IRAS mission, when a large number of low-resolution
spectra of AGB stars was taken, it became clear that the silicate
feature at 9.7 $\mu$m is the more prominent band in the mid-infrared
in most cases. Spectra were classified as having this feature in
emission (type 2n), or in absorption (3n). Sources showing the SiC
feature at 11.3 $\mu$m, which is characteristic of carbon-rich AGB
stars, are classified as type 4n. This feature is much less intense
than the silicate one. As a result, the distribution of energy of
carbon-rich stars in the mid-infrared is close to blackbody.
On the other hand, the silicate feature at 9.7 $\mu m$ can affect the
IRAS [12] magnitude significantly.

Epchtein et al. (1987) studied a sample containing different kinds of
stars found during the ``Valinhos'' survey and studied how the IRAS
magnitudes are affected by the SiC and SiO features. According to the authors,
stars classified in the IRAS-LRS catalogue as 4n are all distributed in a
region in the (K$-$L)$\times$[12]$-$[25] which is close to the blackbody
line. In this diagram, the (K$-$L) colour is interpreted as a sequence
of increasing mass-loss rate or optical depth of the circumstellar dust
shell. The silicate feature at $9.7 \mu$m can appear either in emission or
absorption, depending on the optical depth of the envelope. It is
located well within the IRAS-[12] and the MSX-A band ranges.
On the other hand, a much weaker and broader silicate feature exists at
$18 \mu$m, which generally appears in emission. This feature affects
little the IRAS-[25] magnitude, whereas the MSX photometric bands are
not affected at all. Consequently, the [12]$-$[25] colour index depends
strongly on the strength of the silicate dust feature at $9.7 \mu$m.
When in emission (type 2n sources), the 9.7$\mu$m feature {\it increases}
the star's flux in the [12] band, which makes this magnitude smaller.
On the other hand, when in absorption (type 3n sources), this feature
causes the [12] magnitude to increase. Although the [25] magnitude may
also change with optical depth, the [12]$-$[25] colour index is expected
to be larger for type 3n stars, which are those that exhibit larger
mass-loss rates (the OH/IR stars). In short, they found that O-rich stars
appear separated from C-rich stars in the (K$-$L)$\times$[12]$-$[25]
diagram, and this is caused by the strength of the silicate feature at
$9.7 \mu$m.

Besides Epchtein et al., many papers suggest colour-colour diagrams
as a tool to discriminate carbon from oxygen-rich stars. For instance,
Van Loon et al. (1997, 1998) used the K$-$[12]$\times$ (H$-$K) colour
diagram and van der Veen \& Habing (1988) placed carbon stars in sector
VIa and VII of their [12]$-$[25]$\times$[25]$-$[60] colour-colour diagram.
Whatever the colour-colour diagram used, it generally contains the [12]
band or any other band that includes the 9.7 $\mu$m feature, as the MSX
A band does. In the following, we will discuss where evolved objects are
located in colour-colour diagrams.

Figure 2 shows a colour-colour diagram that is suitable for discriminating
carbon stars from other evolved objects. The lower dotted line limits
region A, where carbon stars are found and oxygen-rich stars are absent.
The former appear to follow the blackbody line down to $T \simeq 900$
K. OH/IR stars are distinguished from carbon-rich objects especially by
the [8.3]$-$[14.7] colour index, as discussed before in this section,
consequently they are found above the carbon stars in this diagram.
Objects which show lower mass-loss rates are distributed
predominantly in region B, while those that have higher values are mostly
in region C, which corresponds approximately to the upper-right region
of Fig. 3, as we will eventually discuss.

Regions B and C are separated by a gap that runs along the
upper dotted line in Fig. 2. This gap was formerly reported by
Sevenster (2002a) in her [8]$-$[12] and [15]$-$[21] histograms and was
interpreted as a separation between late-OH/IR and transition objects.
As she points out, OH/IR stars are mostly located in the third quadrant
of the [8]$-$[12]$\times$[15]$-$[21] colour-colour diagram, which
corresponds to region B in our Fig. 2, and the lower-left corner in
Fig. 3. According to Sevenster, the redder OH/IR objects are mainly
non-variable, post-AGB stars, which corresponds to region C of our
diagram. The transition objects selected in this paper are found spread
out in the three regions: there are 5, 24 and 27 objects in regions
A, B and C, respectively. The proportion of planetary nebulae in the
three regions follows approximately the same distribution. The figures
shown above suggest that this diagram alone does not provide a clear
discrimination between the redder OH/IR objects, transition objects,
and planetary nebulae, but it does provide a fine distinction between
C-rich and OH/IR stars.

We looked for information on the five transition objects that are located
in region A, and we found that two of them are confirmed carbon-rich
stars, two stars seem to be oxygen-rich, and one is undetermined.
MSX5C G290.0075+01.2915 = IRAS 11064-5842 = V382 Car is a yellow (G4 0-I)
supergiant which is listed in the list of Oudmaijer et al. (1992) as
a star showing strong infrared excess. Its infrared spectrum shows several
features between 6 and 9 micron, characteristic of
stretching and bending vibrations of polycyclic aromatic hydrocarbons
(PAH).
MSX5CG 012.2187+04.9223 = IRAS 17514-1555 was suggested to be a transition
object by Hu et al. (1993), and has its optical spectrum classified as WC 11
(Pena, Medina \& Stasinska 2003). The two oxygen-rich objects are
MSX5CG 240.6415+03.0726 (= IRAS 07547-2244 = 11 Pup) and
MSX5CG 015.3218-04.2673 (= IRAS 18313-1738 = MWC939).
The former one has F7/F8 II spectral type and its C/O abundance ratio is
0.28 (Luck \& Webfer 1995). It was extracted from the list of supergiant
stars with infrared excesses compiled by Oudmaijer.
The second oxygen-rich star is of Be type (Vijapurkar et al. 1998)
and shows the silicate feature at $\lambda = 9.8 \mu$m, of the SiO.
The fifth object, MSX5CG 292.9528+01.8774 (= $o$ Cen), for which we did
not find information on its C/O ratio, is a G0 Ia supergiant with
infrared excess (Oudmaijer et al. 1992). These five objects have been
selected by their optical characteristics (low gravity, early or intermediate
spectral type), some of them associated with peculiar infrared properties
(infrared excess). Their location in the K$-$[8.3]$\times$[8.3]$-$[14.7]
colour-colour diagram near the blackbody line indicates that they have
no significant [8.3]$-$[14.7] infrared excess, else they would be in
regions B or C. We must recall that the stars extracted from the list of
Oudmaijer were selected according to the criteria:
IRAS [12]$-$[25]$>0.4$ and [25]$-$[60]$>0.3$, which might result in stars
with colour temperatures in some cases as high as $\sim 1000$ K. On the other
hand, the transition objects located in regions B and C have much lower
colour temperatures. As we discuss in Sect. 5, the tracks that represent
the stellar evolution beyond the AGB phase show that the MSX colour
temperatures remain low long after the star leaves the AGB. Therefore
it is possible that the few transition objects found in region A are not
transition objects, even though they are considered ``suspect''
in the literature.

The behaviour of extinction in the MSX wavelength range is a little
controversial, since it strongly depends on the chemical composition
of the absorbing matter. In this work we consider the extinction curves
given by Cox (2000) and Li \& Greenberg (1997), and we have adopted their
average, as follows:
$A_A/A_V=0.024$ ($\lambda= 8.3\mu$m); $A_C/A_V=0.022$
($\lambda =12.1\mu$m); $A_D/A_V=0.018$ ($\lambda =14.7\mu$m);
and $A_E/A_V=0.023$ ($\lambda= 21.3\mu$m).
It is interesting to observe that the [14.7]$-$[21.3] colour becomes bluer
with increasing extinction. Lumsden et al.(2002) found similar results,
based on other references to extinction curves. Considering the discussion
above, and the fact that the objects in this study are mostly a few
kiloparsecs away, we can state that all {\it pure} MSX colours are
little affected by reddening. The K$-$[8.3] colour is just 0.66 magnitude
reddened for $A_V=10$ and is the most affected colour in this study
(see Fig. 2), however this does not affect the classification of
the sources in the regions A, B, and C, since the reddening vector
is almost parallel to the lines that separate these regions.

\begin{figure}
\includegraphics[width=9cm]{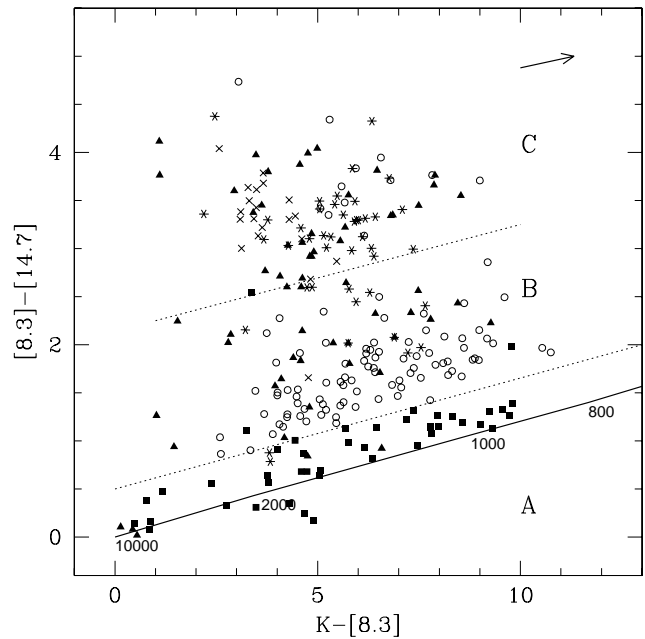}
\caption[]{MSX$-$K colour-colour diagram showing carbon-rich stars,
OH/IR objects, transition objects, and planetary nebulae. The symbols
represent: carbon-rich stars (squares), OH/IR objects (open circles),
transition objects (triangles), and planetary nebulae (asterisks for
well determined K measurements and crosses for lower limits of K).
The continuous line is the blackbody temperature sequence, and the two
dotted lines separate regions A, B and C in the diagram. The arrow in
the upper right corner represents the reddening for $A_V=20$ magnitudes.
}
\end{figure}

Figure 3 shows the various classes of evolved objects in the
[8.3]$-$[14.7]$\times$[14.7]$-$[21.3] colour-colour diagram. This diagram
corresponds approximately to the [8]$-$[12]$\times$[15]$-$[21] of
Sevenster (2002a), her Fig. 3.
The blackbody line is drawn in, together with the corresponding colour
temperatures. We can define a region in the diagram where carbon
stars are located, which is near the blackbody line at $T=800 \sim 2000$
K. These stars show a range of near-infrared colours as wide as
OH/IR objects do (see Fig. 1); however, at mid-infrared wavelengths
the colour temperatures of carbon stars are higher.
Near the region occupied by C-stars, but in a slightly lower temperature
range ($T=800 \sim 1000$ K), a few OH/IR objects are also observed; however,
the majority of these shows temperatures in the range $T=500 \sim 800$K.
As we shall discuss in Sect. 5, these OH/IR objects probably are
oxygen-rich AGB stars.
A second, distinct group of OH/IR objects is spread over a large area in
the diagram, to the upper right of the dotted line, together with
transition objects and planetary nebulae. The latter group of objects
is separated from the first by a gap, which is indicated in Fig. 3
by the dotted line; it corresponds to the dips seen in the MSX colour
histograms drawn by Sevenster. Because the definitions of magnitudes used
in this paper are different from those used by Sevenster the gaps do not
have the same colour indices as the dips. In this diagram, the OH/IR
objects to the upper right of the dotted line fall in the same region as
transition objects and planetary nebulae. These OH/IR objects might be
either stars close to leaving the AGB or transition objects with oxygen-rich
envelopes.

The gap that runs along the dotted line in Fig. 3 is not a bias effect,
but results from the rapid evolution of the stars as they cross it. Below
we discuss this statement, which will be strengthened in Sect. 5, where
we discuss results of theoretical models. As we remark in Sect. 2,
the samples of OH/IR objects and C-rich stars both contain objects in a
wide range of colour indices, which are indicative of a large variety of
mass-loss rates and optical depths as well. These two parameters also depend
also on the evolutionary stage of the star in the AGB phase. L\'epine et al.
(1995) and Epchtein et al. (1990) showed that OH/IR and C-rich stars
in the various evolutionary stages are smoothly distributed along sequences
in near-infrared colour-colour diagrams.
Therefore, since the stars in our samples are also smoothly distributed
in colour (Fig. 1), we conclude that these two samples contain stars
smoothly distributed concerning mass-loss rates, optical depths of
the envelopes, and evolutionary stages.

At this point, a word must be said on the evolution of carbon-rich AGB
stars in the colour-colour diagrams discussed above. As one can see in
Fig. 3, there are no carbon-rich stars on the upper right side of the
dotted line, where numerous OH/IR objects coincide with transition
objects and planetary nebulae. This absence deserves an explanation, since
carbon-rich stars are the result of the evolution of oxygen-rich AGB
stars in the thermal-pulse phase, where the abundance of carbon exceeds
that of oxygen. Therefore, it seems that carbon-rich AGB stars should
coincide with OH/IR stars: both classes show high mass-loss rates,
they are both in the thermal-pulse phase and have optically thick
envelopes. However, even the redder carbon-rich stars have higher
colour temperatures than OH/IR stars. Here we discuss the reasons for
this discrepancy. Ordinary carbon-rich stars have their spectral energy
distribution (SED) maximum near $1-2\mu$m. Despite the maximum shifting to
longer wavelengths in the case of very red carbon-rich stars (like
IRC+10216, for example), its position is still located
at wavelengths near to or shorter than 8 $\mu$m. The carbon-rich stars that
one would expect to find in the upper right corner of MSX colour-colour
diagrams are the {\it extreme} carbon stars (e.g. AFGL 3068), which have
maxima near $\lambda = 10 \mu$m. However, even these extreme examples
have moderate colour temperatures, about $T = 2000 - 2500$ K. Thus, even
the extreme carbon stars have temperatures which put these stars at the
left bottom corner of Fig. 3. On the other hand, carbon-rich
{\it transition objects} can show large MSX colour indices, because
their SEDs are double peaked: they exhibit a maximum near $\lambda =
1 \mu$m , which originates from the central star, and a second (broad)
maximum near $\lambda = 10 - 20 \mu$m. This second maximum accounts for
the low MSX colour temperatures of these objects.

Both colour-colour diagrams discussed above were formerly proposed by
Lumsden et al. (2002) in a paper where various types of young objects such
as compact HII regions, massive young stellar objects, Herbig Ae/Be stars,
are also discussed. Most of the results found in this work confirm
their previous conclusions, but here we outline a few differences.
As one can see in Fig. 1, the sample of carbon-rich stars contains
a significant number of objects with optically thick circumstellar
envelopes, whereas there are few of them in the work of Lumsden
and his colleagues. The sample of carbon stars used in this work is
based on the compilation by Loup et al. (1993), which is a collection
of radio observations of the CO and HCN molecules. This list does not
have an {\it optical} bias, which would select only stars with optically
thin envelopes. On the other hand, Lumsden et al.(2002) remark that most of
the carbon stars used in their work have been optically selected,
although they include in their study a few heavily embedded carbon stars.
Another important difference between our [14.7]$-$[21.3]
$\times$[8.3]$-$[14.7] diagram and Lumsden's is the optical depth range
in the sample of OH/IR objects. Our sample, which is the same as used by
Sevenster (2002a,b), contains a large number of OH/IR objects sharing the
same space in the colour-colour diagram with planetary nebulae and
transition objects. These OH/IR objects are about to leave the AGB branch,
as we discuss in the next section and as has been reported by Sevenster
(2002a,b) and they are missing in the work of Lumsden.

\begin{figure}
\includegraphics[width=9cm]{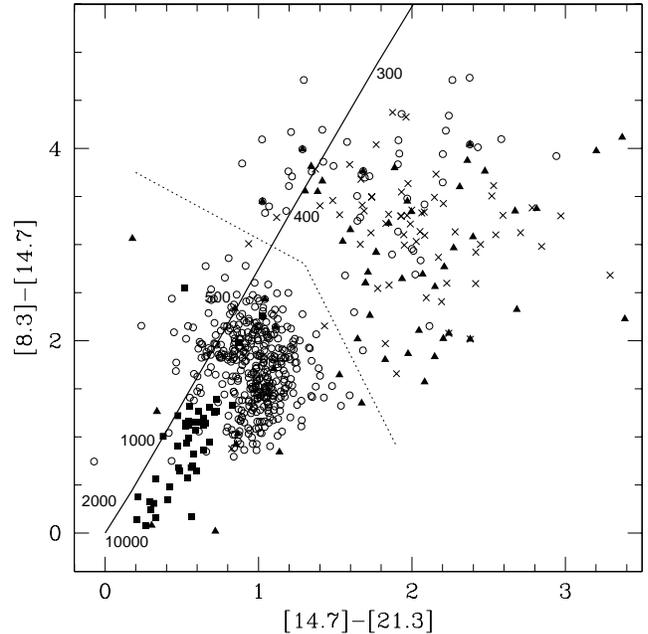}
\caption[]{MSX colour-colour diagram for carbon-rich stars, OH/IR objects,
transition objects, and planetary nebulae. The symbols used are the same
as those in Fig. 2 and the continuous line represents the blackbody
temperature sequence.}
\end{figure}

\section{Classification of very red OH/IR objects}

In Sect. 3 we identified a population of OH/IR objects with high
colour indices that are separated from the main bulk of OH/IR objects
by a gap (Fig. 3). Hereafter those objects will be referred as
``very red OH/IR objects'' (VR-OH/IR objects). Sevenster (2002a) used
MSX and IRAS colour indices to study her sample of OH/IR objects and
found a similar sub-sample of very red OH/IR objects in her
[8]$-$[12]$\times$[15]$-$[21] colour-colour diagram. A comparison
between that diagram and Fig. 3 of the present work reveals that the
VR-OH/IR objects correspond mostly to the OH/IR objects in quadrants
I and IV of Sevenster. This correspondence is straightforward because
the colours [8]$-$[12] and [8.3]$-$[14.7] are quite similar. Sevenster
also carried out a study of the OH spectra of the various types of OH/IR
object, concerning their position in MSX and IRAS colour-colour diagrams.
She points out that the majority (61 to 80\%) of the OH/IR objects are
double-peaked, and the fraction of OH/IR objects that show single-peak
or irregular OH spectra does not change significantly with the mid-infrared
colours. On the other hand, the expansion velocity of the OH envelope is
related to the [8]$-$[12] colour index in the sense that the VR-OH/IR
objects rarely show expansion velocities greater than 15 km/s.
Since some of these objects might be post-AGB stars, she concludes
also that the outflow velocities of the OH envelope of post-AGB stars are
lower than 15 km/s. In spite of the conclusions drawn above, the nature
of the members of the group of VR-OH/IR stars remained ``possibly
post-AGB'' or ``late-AGB oxygen-rich stars'', whilst the relative
contribution of the various classes is still undetermined.
The aim of this section is to present the results of a survey in the
literature intended to establish the nature of the VR-OH/IR objects.

The VR-OH/IR objects constitute a mixed population,
which includes objects that are still in the AGB (the OH/IR stars with
very high mass-loss rates), sources that have already left this stage
(transition objects, planetary nebulae), and young objects
(protostars, molecular clouds, etc.). We used the SIMBAD database to
find literature on all 49 VR-OH/IR objects. For each object we
examined the original references for a classification other than
``OH/IR object'' or ``infrared source''. Objects classified in the
literature as ``proto-planetary nebula'' and ``post-AGB star'' have been
brought together under the designation ``transition objects''. We have kept
the original term ``possible PN'' for a possible planetary nebula and
``HII'' for a HII region, which might be interpreted as an OH/IR object of
interstellar origin near a HII region. We emphasize that we did not
classify the objects, but simply compiled the classifications given
in the original references. When a source did not have a classification,
it was considered as ``OH source''.

In addition to the classification described above, we examined visually
the 49 OH spectra and classified them according to the following criteria:

\begin{enumerate}
\item 2P $=$ double peak
\item 2Pa1 $=$ asymmetric double peak, low velocity weaker
\item 2Pa2 $=$ asymmetric double peak, high velocity weaker
\item 1P $=$ single peak
\item B $=$ bump emission (one single, broad peak)
\end{enumerate}

\noindent We also include the L classification that can be added to
the types above for a broad line (except the B type). We consider
a {\it broad} feature (L) the cases where the full width at half
maximum (FWHM) in units of km/s is in the range $15 < FWHM < 25$
and a {\it bump} (B) when $FWHM > 25$. 

Table 5 shows the OH names of the VR-OH/IR sources, the type of OH
spectrum, and its classification according to the original references
given in the SIMBAD database. Among the 49 VR-OH/IR objects, 2 (4\%) are
classified as planetary nebulae, 2 (4\%) are cited as possible
planetary nebulae, 20 (41\%) as transition objects, 9 (18\%) as OH/IR stars,
14 (29\%) as OH sources (i.e. without classification), and 2 objects
(4\%) are associated with HII regions. The three former categories
(49\% of the objects) comprise objects that have already left the AGB,
but this figure might be even larger, since some of the 14 objects cited
as OH sources lack a more detailed examination and might be in the post-AGB
evolutionary phase. Finally, a minor part of the objects is believed to
be related to star forming regions, which is in agreement with the results of
Felli et al. (2000), who identified young stellar objects in the ISOGAL
survey with similar colours ([7]$-$[15]$>2.5$) in a field near $l=+45^o$.

An examination of the OH types of the various classes of object in Table
5 confirms the statement by Sevenster (2002a) that transition
objects can be found with different kinds of OH spectra, i.e. the OH
spectrum alone does not determine whether an object has left the AGB
or not. In fact, the transition objects cited in the literature
might have double-peak (18 objects) or single-peak (2 objects) OH spectra.

\begin{table*}
\caption[]{Very red OH/IR sources; the OH type is defined in the text.
The classification is given as in the original reference.
The OH references are: A= Sevenster et al. (1997a), B= Sevenster et al.
(1997b), and C= Sevenster et al. (2001).
The references between brackets are: (a)= Nyman et al. (1998), (b)=
van den Steene et al. (2000), (c)= Hrivnak et al. (1999), (d)= 
Preite-Martinez (1988), (e)= Gomez et al. (1990), (f)= Deguchi et al.
(2000), (g)= Garcia-Lario et al. (1997), (h)= Fish et al. (2003),
(i)= Engels (2002), (j)= Zijlstra et al. (1989), (k)= Likkel et al.
(1991), (l)= Likkel (1989), (m)= Kohoutek (2002), (n)= Walsh et al.
(1998); (o)=Sevenster (2002b); (p)=te Lintel Hekkert et al. (1989);
(q)=Zijstra et al. (2001); (r)=Blommaert et al. (1994); (s)=Lewis et
al. (1990); and (t)=Fix \& Mutel (1984).
}
\begin{tabular}{|c|c|c|l|l|l|}
\hline
\hline
\noalign{\smallskip}
 OH name           & OH type & OH Ref. & Classification & other masers &
Refs. \\
\noalign{\smallskip}
\hline
 OH 314.933-02.052 &  2Pal & B &  Possible PN             & SiO & (a) \\
 OH 323.459-00.079 &  1P   & B &  HII region              & & (n) \\
 OH 326.932+02.754 &  2P   & B &  OH source               & & \\
 OH 326.530-00.419 &  2Pa1-L & B & OH source              & & \\
 OH 335.832+01.434 &  2P   & B &  Post-AGB Star           & & (b,o) \\
 OH 337.065-01.173 &  1P   & B &  OH source               & & \\
 OH 353.844+02.984 &  1P   & A &  "Cotton Candy" - Post-AGB  & & \\
 OH 353.973+02.727 &  2Pa2 & B &  PN (GLMP 546)           & & (m,o) \\
 OH 348.813-02.840 &  2Pa1 & B &  "Walnut Nebulae" - Post-AGB & & (c)\\
 OH 353.945-00.972 &  2Pa1 & A &  Post-AGB Star           & & (d,o)\\
 OH 359.140+01.137 &  2Pa1 & A &  Post-AGB Star           & H$_2$O, SiO &
(e,o)\\
 OH 002.348+02.965 &  2P   & A &  OH/IR Star              & SiO & (f) \\
 OH 357.092-00.362 &  2Pa2 & A &  OH/IR Star              & & (p) \\
 OH 000.207+01.414 &  1P-L & A &  IRAS 17375-2759  PN     & SiO & (a) \\
 OH 359.233-01.876 &  2P   & A &  Post-AGB Star           & & (g,o) \\
 OH 002.286-01.801 &  2Pa1 & A &  Post-AGB Star           & & (g,o) \\
 OH 005.885-00.392 &  1P   & C &  HII region              & & (h) \\
 OH 003.471-01.853 &  B    & A &  IRAS 17576-2653 - Possible PN & & (d)\\
 OH 004.017-01.679 &  2P   & A &  Post-AGB Star           & & (g,o) \\
 OH 011.560+00.088 &  2Pa1 & C &  OH/IR Star              & & (j) \\
 OH 015.364+01.925 &  2Pa2 & C &  Post-AGB Star           & & (g,o) \\
 OH 016.240-00.622 &  2Pa1 & C &  OH source               & & \\
 OH 020.854+00.486 &  B    & C &  OH/IR Star              & & (p,q) \\
 OH 017.684-02.032 &  2P   & C &  Post-AGB Star           & H$_2$O & (i,o) \\
 OH 023.751+00.210 &  2Pa1 & C &  OH/IR Star              & & (j) \\
 OH 025.154+00.060 &  2P   & C &  OH/IR Star              & & (r)\\
 OH 026.063-00.519 &  2Pa1 & C &  OH source               & & \\
 OH 027.577-00.853 &  2P   & C &  Post-AGB Star           & & (g,o) \\
 OH 038.909+03.178 &  2Pa1 & C &  Post-AGB Star           & &  (g,o) \\
 OH 038.287+01.875 &  2Pa1 & C   &  OH/IR Star            & & (s)\\
 OH 035.209-02.653 &  2P   & C   &  GLMP 870 - Post-AGB Star  & H$_2$O &
(g,k,l)\\
 OH 043.165-00.028  &  1P   & C   & OH source              & & \\
 OH 341.275-00.720  &  2Pa2 &  B  & OH source              & & \\
 OH 345.052-01.855  &  2Pa1 &  B  & OH source              & & \\
 OH 348.668-00.715  &  2Pa2 &  B  & OH source              & & \\
 OH 349.804-00.321  &  2Pa2 &  B  & Post-AGB Star          & & (g,o) \\
 OH 351.474-00.596  &  2Pa2 &  A  & OH source              & & \\
 OH 353.637+00.815  &  2Pa2 &  A  & Post-AGB Star          & & (o) \\
 OH 000.260+01.027  &  2Pa2 &  A  & OH source              & & \\
 OH 001.484-00.061  &  2Pa1 &  A  & OH/IR Star             & & (p) \\
 OH 007.961+01.445  &  2P   &  C  & Post-AGB Star          & & (g,o) \\
 OH 009.097-00.392  &  1P   &  C  & OH source              & & \\
 OH 006.594-02.011  &  2Pa2 &  C  & Post-AGB Star          & & (o) \\
 OH 007.420-02.042  &  2Pa1 &  C  & OH source              & & \\
 OH 015.700+00.770  &  2Pa2 &  C  & Post-AGB Star          & H$_2$O  & (i,o) \\
 OH 022.043-00.608  &  2Pa2 &  C  & OH/IR Star             & & (t)\\
 OH 030.944+00.035  &  1P   &  C  & OH source              & & \\
 OH 030.394-00.706  &  1P   &  C  & Post-AGB Star          & & (o) \\
 OH 037.118-00.847  &  2Pa2 &  C  & Post-AGB Star          & H$_2$O   & (i,o)\\

\noalign{\smallskip}
\hline
\hline
\end{tabular}
\end{table*}

\section{Theoretical tracks beyond the AGB phase}

In this section, we compare the colour-colour diagrams described in the
previous section with theoretical predictions of magnitudes and colours.
The model for post-AGB evolution assumed in this work was proposed in
Volk (1993) and is shortly described here.

Each model starts when the star leaves the AGB i.e., its envelope becomes
detached from the core and is driven away at the speed of 10 km/s. The
evolution of the core is that calculated by Sch\"onberner (1981, 1983) for
$0.64 M_{\odot}$, but accelerated by a factor 2 in order to match the
IRAS colours of planetary nebulae. It is well known that the evolution
time in the post-AGB phase depends on the mass-loss rate during the AGB,
and authors have found a wide range of times, depending on the mass-loss
rate assumed. Another hypothesis assumed in the models used is that the
star leaves the AGB burning hydrogen, i.e. {\it between} two thermal
pulses. According to Iben (1984), post-AGB evolution is dominated by
helium burning if the star leaves the AGB in the phase range
$0 < \phi < 0.15$, where the phase $\phi = 0$ at the time of the
beginning of the thermal pulse. Post-AGB stars might also have a
{\it late pulse} during the horizontal evolution across the HR diagram,
if the star leaves the AGB near to having a pulse, let us say for $\phi > 0.9$,
approximately. During the post-AGB helium burning the star loops in the
HR diagram and returns to the AGB in about $200$ years (Bl\"ocker 1995).
Considering the figures above, we can state that helium burning applies
in the minority of the cases. As a result, the effects of thermal pulses
for stellar evolution in colour-colour diagrams are presently unknown and
difficult to predict.

The circumstellar dust shell and the neutral gas are modelled
following Spagna \& Leung (1983). The models end at 10,000 years of
evolution, when numerous emission lines dominate the spectrum together
with a strong infrared continuum in the mid-infrared. The mass-loss rates
at the end of the AGB are $2.1 \times 10^{-5}M_{\odot}$/year for models
{\it a} and {\it c} and $5.2 \times 10^{-6}M_{\odot}$/year for models
{\it b} and {\it d}, respectively.
We take four different sets of parameters for the circumstellar dust
shell in Volk's model: (a) silicates, initial optical depth 10.0 at 9.7
$\mu$m; (b) silicates, initial optical depth 2.5 at 9.7 $\mu$m; (c)
graphite, initial optical depth 2.5 at 11.22 $\mu$m; (d) graphite,
initial optical depth 0.625 at 11.22 $\mu$m.
Each set of spectra (two silicate and two graphite) was convolved with
the MSX filter curves given in the MSX Explanatory Guide (Egan et al.,
1999). We do not consider colour-colour diagrams that include the K band
because of limitations in the program that cause this magnitude to be
inaccurate.

Figures 4 and 5 reproduce the [14.7]$-$[21.3]$\times$[8.3]$-$[14.7]
described in the previous section together with the evolutionary tracks
described above. As one can see in
the figures, the four evolutionary tracks start at the crowd of points that
represent AGB stars with high mass-loss rates. The time the star spends
together with the AGB stars is extremely short: less than 100 and 300
years for oxygen and carbon-rich stars, respectively. This short time
accounts for the gap that separates the VR-OH/IR objects, planetary
nebulae, and transition objects, from the bluer group of OH/IR objects
(dotted line in Figs. 3, 4, and 5). Eventually the stars evolve rapidly
towards the upper-right part of the diagram, into the region where
transition objects, planetary nebulae, and VR-OH/IR objects are located.
In this region the tracks are quite clumsy but they cover the cloud of points
that are spread over a wide range of colours. Here some differences between
oxygen- and carbon-rich objects should be noted: according to the models,
O-rich objects can have a redder colour after 400 years of post-AGB evolution,
whilst the carbon-rich ones are more restricted in colour and are not
expected to be redder than [14.7]$-$[21.3]$=$2.2. Eventually the star will
leave the region where transition objects and PNe are, and head for the
upper-left region of the diagram in a time that will strongly depend on
the parameters of the model: 2200 years for models {\it a} and {\it c}
(that have higher mass-loss rates), and 1600 years for models {\it b} and
{\it d}, which corresponds to core temperatures of $1.9 \times 10^5$ K and
$1.1 \times 10^5$ K, respectively.
After that point, models and data disagree completely. The diagrams show
that transition objects and planetary nebulae should have similar colours
but the models predict that planetary nebulae evolve differently, first
having high [8.3]$-$[14.7] colour index and eventually having negative
[14.7]$-$[21.3] colour index. Our sample contains planetary nebulae in
various evolutionary stages, but there are no PNe with colours that match
the predictions of the models after 2200 years of evolution, therefore
we can conclude that they do not work correctly after that point.
Volk (1992) used another grid of models to study planetary nebulae in
the IRAS [12]$-$[25]$\times$[25]$-$[60] colour-colour diagram and also in
his diagrams the evolutionary tracks reproduce well the colours of PNe,
but they end up quite far from the main concentration of points at $t \simeq
7 \times 10^3$ years of evolution. These differences of $t$ between the
various models occur because how quickly a post-AGB evolves depends mainly
on the mass-loss rate assumed in the AGB and this has been chosen to
match the observations. Apart from these discrepancies, the models show
that the wide range of colours observed is not caused by variability or
bad measurements, although they can play a role, but is the
result of normal evolution during the post-AGB phase. The models also
explain why there is a gap between the crowd of points representing
late-AGB stars and transition objects: stars cross that gap very rapidly,
in two hundred years or less.

\begin{figure}
\includegraphics[width=9cm]{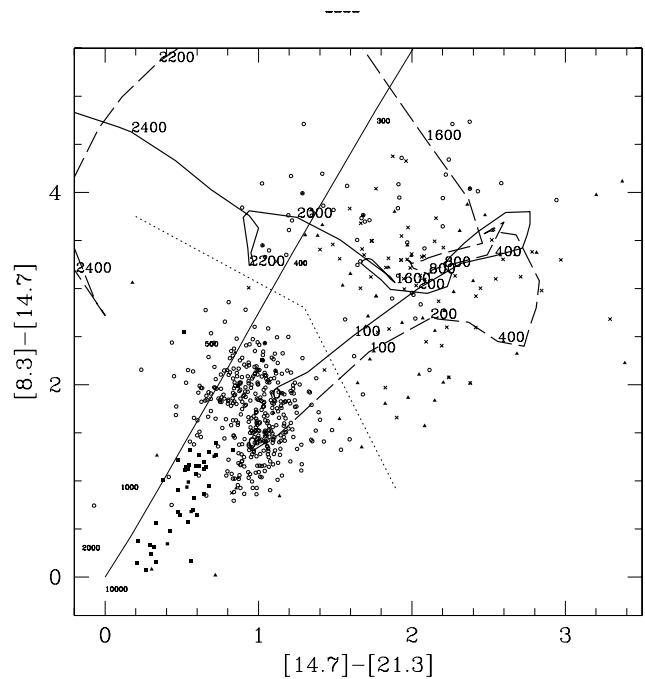}
\caption[]{Evolutionary tracks of an oxygen-rich star in the post-AGB phase.
Evolution time is plotted next to the curves and is counted in years.
The two curves correspond respectively to mass loss rates of $2.1 \times
10^{-5} M_{\odot }$/year (model a, continuous line) and $5.2 \times
10^{-6} M_{\odot }$/year (model b, dashed line).
The symbols are the same as in Figs. 2 and 3 but have been shrunk
to make the curves clearer.}
\end{figure}

\begin{figure}
\includegraphics[width=9cm]{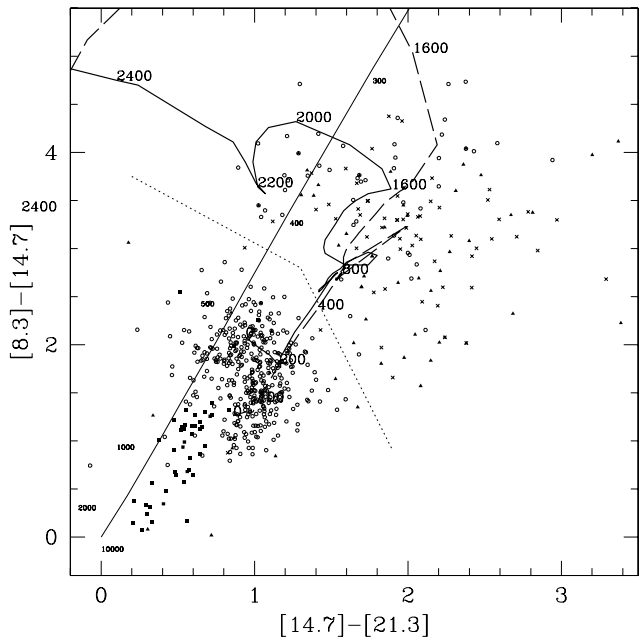}
\caption[]{The same as Fig. 4, but for carbon-rich stars with mass-loss
rates of $2.1 \times 10^{-5} M_{\odot }$/year (model c, continuous line)
and $5.2 \times 10^{-6} M_{\odot }$/year (model d, dashed line).
Symbols used are the same as in Figs. 2 and 3.}
\end{figure}

Sevenster (2002b) presents evolutionary tracks on colour-colour diagrams
for the three IRAS bands and the K band, based on the models proposed by
van Hoof et al. (1997). Those models do not provide MSX magnitudes and
colours, therefore they cannot be quantitatively compared with our results.
Nevertheless there is a feature that is remarkable in their model, but is
absent in ours: the red loop at the end of the AGB phase, which causes
the $R_{21}$ colour to loop back towards the blue, before the star
resumes its normal evolution to the red. Evolved objects are scarcely
found along the blue part of those tracks, but there is a significant
group of objects along the red part. The models provided
by Volk used in this work do not predict {\it red loops}, and they
seem to fit the data better. According to the present model, evolution
towards blue colours is predicted to happen only after about $2 \times
10^3$ years of post-AGB evolution (see Figs. 4 and 5), when the star
is halfway to the planetary nebula phase.

\section{Conclusions}

We compiled a sample containing various classes of evolved objects, such
as: carbon-stars, OH/IR objects, transition (post-AGB) objects, and
planetary nebulae. We searched for mid and near-infrared counterparts
in the MSX survey and in the literature, and constructed
colour-colour diagrams to study the evolution of these objects.
We found that oxygen-rich, AGB stars (OH/IR) can be distinguished from
carbon-rich stars in the (K$-$[8.3])$\times$([8.3]$-$[14.7]) colour-colour
diagram, similarly to what was proposed before by Epchtein et al.
(1987). We explain the separation between the two groups of stars as the
impact of the SiO $9.7\mu$m feature on the [8.3] MSX band, that shifts
OH/IR stars away from the blackbody line. Transition objects, high
mass-loss OH/IR objects, and planetary nebulae occupy the same region in
this colour-colour diagram and in the ([14.7]$-$[21.3])$\times$([8.3]$-$[14.7])
diagram as well. We confirm Sevenster's (2002a) previous finding that
the majority of OH spectra of transition objects are normal, double-peaked.
Nevertheless, among the 49 objects that show evidence of high mass-loss
rate, a minor but significant percentage (22\%) has OH spectra that
differ from the classic double peak, perhaps because of some mechanism
related to the superwind phenomenon. We tested a grid of four models
with different C/O ratios, mass-loss rates and optical depths developed
by Volk (1992, 1993). A comparison between the infrared colours indices
predicted by the model and the observations showed that the gap that
separates objects with optically thick envelopes from those with
small optical depth values is caused by the rapid evolution just after
the star leaves the AGB. Eventually, the objects describe a tortuous
trajectory in MSX colour-colour diagrams, which explains the wide range
of colour indices of transition objects and planetary nebulae.
The models reproduce well the infrared MSX colours up to $\sim 2000$
years after the end of the AGB phase, i.e. when the temperature of the
core reaches $1 - 2 \times 10^5$ K.

\begin{acknowledgements}

We are indebted to Kevin Volk for providing us his models of post-AGB
evolution. We thank the anonymous referee for his valuable suggestions
and critical reading of the manuscript. This publication makes use of
data products from the Two Micron All Sky Survey, which is a joint
project of the University of Massachusetts and the Infrared Processing
and Analysis Center, funded by the National Aeronautics and Space
Administration and the National Science Foundation. This research has
made use of the SIMBAD database, operated at CDS, Strasbourg, France.
This work was supported by FAPESP (Funda\c c\~ao de Amparo \`a
Pesquisa do Estado de S\~ao Paulo, Brazil) under grant number
2002/08816-5, and CNPq (Conselho Nacional de Desenvolvimento Cient\'\i fico
e Tecnol\'ogico, Brazil). S.L.M. and W.J.M. acknowledge CNPq
(processes 478309/01-5 and 3001072003-0) for financial support.

\end{acknowledgements}

\end{document}